\documentclass[aps,showpacs]{revtex4}
\usepackage{epsfig}
\begin{document}

\title{Pionic atoms probing $\pi NN$ resonances}
\author{
M. I. Krivoruchenko$^{*, \dagger )}$, B. V. Martemyanov$^{*, \dagger )}$,
Amand Faessler$^{\dagger )}$ and C. Fuchs$^{\dagger )}$ 
}

\affiliation{
$^{*)}$Institute for Theoretical and Experimental Physics$\mathrm{,}$
B. Cheremushkinskaya 25\\ 117259 Moscow, Russia \\
$^{\dagger )}$Institut f\"{u}r Theoretische Physik$\mathrm{,}$ Universit\"{a}t 
T\"{u}bingen$\mathrm{,}$
Auf der Morgenstelle 14\\ D-72076 T\"{u}bingen, Germany
}

\begin{abstract}
The pion optical potential generated by the hypothetical $\pi NN$-coupled $NN$-decoupled dibaryon resonance 
$d^{\prime}(2065)$ is calculated to the lowest order in nuclear matter density. The contribution to the pion 
optical potential is found to be within the empirical errors, so the $d^{\prime}(2065)$ existence currently 
does not contradict to the observed properties of the $\pi^{-}$-nucleus bound states. Future progress in the 
pionic X-ray spectroscopy can reveal contributions of $\pi NN$ resonances to energy levels and widths of the 
pionic atoms.
\end{abstract}

\maketitle

The possibility for existence of dibaryon resonances is discussed over two decades \cite{JAFF,MULD,KON,FILK,SAKA}. 
The most popular candidates for exceptionally narrow dibaryon resonances are the $H$ dihyperon with a mass 
below the $\Lambda \Lambda$ threshold, as predicted by Jaffe \cite{JAFF} within the framework of the MIT bag model, and the nonstrange $\pi NN$-coupled 
$NN$-decoupled $d^{\prime}$ dibaryon with quantum numbers $I=0$ and $J^P=0^{-}$ and a mass close to the 
$\pi NN$ threshold \cite{MULD,KON}. The resonance-like shape of the DCE cross section \cite{BILG} 
was interpreted by Martemyanov and Schepkin \cite{MASCH} as a manifestation of the $d^{\prime}$ dibaryon with 
a mass of $2065$ MeV. Searches of 
the $H$ dihyperon did not give conclusive results
\cite{SAKA,AOKI,BARM96,BELZ,CHRI98}. The upper limits for the $d^{\prime}$ production cross sections are settled experimentally
\cite{SIOD,DORO}, while the most specific features of the DCE reaction have been described using
conventional mechanisms \cite{GIBB,BOBY}. Astrophysical constraints 
to masses and coupling constants of the dibaryons are derived from existence of massive neutron stars \cite{JTPL,PRC,JPG,GLE}.

The laboratory experiments with pionic atoms \cite{KONI,OLIN,LAAT,YAM98,GEI02} provide one more 
opportunity to test existence of the $d^{\prime}$ dibaryon: 
The $d^{\prime}$ is coupled to the $\pi NN$ channel, so as illustrated by Fig.~1 it contributes to the pion optical potential to the second 
order in the nuclear matter density and, as a result, manifests itself in shifts and broadening of the atomic 
spectral lines. 

The pion-nucleus interaction effects on the pion bound states and present experimental status of the pionic X-ray spectroscopy
are reviewed recently by Kienle and Yamazaki \cite{KIEN}.

\begin{figure}[tb]
\par
\begin{center}
\leavevmode
\epsfxsize = 7 cm \epsffile[10 540 380 790]{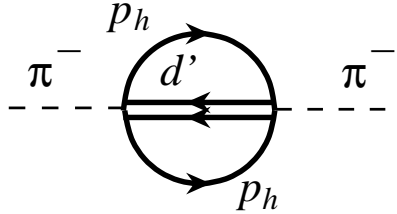}
\end{center}
\caption{
Contribution to the pion optical potential of the $d^{\prime}$ dibaryon (solid double line). The 
single solid lines show the proton holes ($p_h$) in the Fermi sphere, the dashed lines show the pion propagating in the nuclear matter. 
}
\label{fig1}
\end{figure}

The $d^{\prime}$ contribution to the $\pi^{-}$ optical potential at the nuclear density $\rho_0$ has the form
\begin{equation}
2\mu \delta V_{opt}({\bf k}) = \frac{2g^2}{M\eta _F} 
\int{
\frac{d{\bf p}_1}{(2\pi)^3}\frac{d{\bf p}_2}{(2\pi)^3}
\frac{1}{|D_J|^2}
\frac{\theta(p_F - |{\bf p}_1|) \theta(p_F - |{\bf p}_2|)}
{{\bf p}_1^2/(2m) + {\bf p}_2^2/(2m) + {\bf k}^2/(2\mu) - ({\bf p}_1 + {\bf p}_2 + {\bf k})^2/(2M) - \Delta M + i\Gamma/2} 
} \;
\end{equation}
where ${\bf p}_1$ and ${\bf p}_2$ are momenta of two protons, ${\bf k}$ is the pion momentum, 
$m$ is the nucleon mass, 
$\mu$ is the pion mass, $p_F$ is the Fermi momentum, $\Delta M = M - 2m - \mu$, $M = 2065$ MeV and $\Gamma$ are the $d^{\prime}$ 
mass and width, $g = 9.2$ GeV$^{-1}$ is the $\pi NN d^{\prime}$ coupling 
constant determined from the width $\Gamma = 0.5$ MeV of $d^{\prime}$ decay in the vacuum \cite{BILG,SCHE}
\begin{equation}
\Gamma = \frac{3g^2}{64\pi^2}\sqrt{\mu M}\Delta M^2 
\end{equation} 
with enhancement factor $\eta _F$ included to the coupling constant
$g^2$.  
The Jost function $D_J$ in the effective-range approximation is given by \cite{GOLD}
\begin{equation}
D_J = \frac{|{\bf p}_1 - {\bf p}_2|/2 + i\gamma}{|{\bf p}_1 - {\bf p}_2|/2 + i\alpha}
\end{equation}
where $\alpha \approx 2/r_e$ and $\gamma = 2/a/\alpha/r_e$, with $a = 7.82$ fm being the effective proton-proton scattering 
length including the Coulomb interaction \cite{BRJE}, and $r_e = 2.67$ fm the effective radius. The enhancement factor 
$\eta _F \approx 1 + 32/(r_e^2 M\Delta M) \approx 3$ according to \cite{HI}. It is determined as an average value
of the $1/|D_J|^2$ over the three-body $\pi NN$ phase space. 

The $d^\prime$ in-medium width is estimated to be $\Gamma^{*} = 10$ MeV \cite{BILG}. It appears due to the reactions
$d^{\prime}N \rightarrow NNN$. It is of the first order in the density, so formally we go beyond the lowest second-order 
calculation. The effect, however, is numerically large. The $d^{\prime}$ width is increased as compared to the vacuum 
value by a factor of 20. The vacuum decay channel $d^{\prime} \rightarrow \pi NN$ is blocked in our case, since the pion 
is bound. The collision broadening of the $d^{\prime}$ is the only effect contributing to the imaginary parts of the 
$d^{\prime}$ optical potential.

The second-order pion optical potential is related to the non-resonant absorption 
of pions on nucleon pairs \cite{ERIC}. It is parameterized in the form
\begin{equation}
2\mu \delta V_{opt}({\bf k}) = \delta q + \delta p {\bf k}^2 
\end{equation}
where 
\begin{eqnarray}
\delta q &=& - 16\pi (1 + \frac{\mu}{2m})B_0\rho _p \rho _n, \\
\delta p &=&  - 16\pi (1 + \frac{\mu}{2m})^{-1}C_0\rho _p \rho _n
\end{eqnarray}
are the $s$- and $p$-wave parts of the pion optical potential, $\rho_p$ and $\rho_n$ are the proton and neutron densities. 
Eq.(1) determines the $d^{\prime}$ contribution to the parameters $B_0$ and $C_0$. 

The results are summarized and compared to the experiment in Table 1. The empirical data are taken from \cite{FRIE02} for two 
realistic models with a vanishing nucleon-nucleon correlations parameter. The results of fit of the $s$-wave part of the $\pi^{-}$- nucleus
potential \cite{KIEN} are also given. The real $s$-wave part should be of the same magnitude as the imaginary part \cite{GARC,OSET,CMKO,SALC}. 
The result of the standard model $|\mathrm{ReC_0}| \sim 3|\mathrm{ImC_0}|$ is not fully satisfactory. The phenomenological pion-nucleon 
isovector scattering length in the standard model also appears to be overestimated as compared to the free value \cite{KIEN,FRIE02}.
A specific version of the relativistic impulse approximation \cite{BIRB83,GOUD91,BIRB91,AGAL92,BIRB92,FRIE} and the recent WF model 
\cite{WEIS,FRIE} reconcile those problems, at least partially. 

\begin{table}[tbp]
\caption{
Contribution of the $d^{\prime}$ dibaryon to the parameters $B_0$ and $C_0$ of the pion optical potential in the symmetric nuclear matter 
at the saturation density $\rho_0 = 0.17$ fm$^{-3}$ 
and en effective density $\rho_{e} \sim 0.6\rho_0$ where the pion-nucleus interaction has its maximum, as compared to the empirical 
values extracted form the pionic X-ray spectroscopy using the standard (S) model,
Weise-Friedman (WF) model \cite{WEIS,FRIE}, and Kienle-Yamazaki (KYa) model \cite{KIEN}. The 
empirical data are from \cite{FRIE02} and \cite{KIEN}, respectively.
The values in parenthesis refer to the case of in-medium width of $d^{\prime}$.}
\label{labl2}
\begin{center}
\begin{tabular}{ccccc}
\hline\hline
$\mathrm{Model}$ 	& $\mu^{4} \mathrm{ReB_0}$ & $\mu^{4} \mathrm{ImB_0}$ &$\mu^{6} \mathrm{ReC_0}$ & $\mu^{6} \mathrm{ImC_0}$ \\ \hline
S            		& $-0.15 \pm 0.04$ 	& $0.054 \pm 0.002$ & $-0.28 \pm 0.01$ & $0.062 \pm 0.003$  \\ 
WF           		& $-0.07 \pm 0.04$ & $0.053 \pm 0.002$ & $-0.28 \pm 0.01$ & $0.067 \pm 0.003$  \\ 
KYa           		& $-0.02 \pm 0.02$ & $0.047 \pm 0.002$ & $ $ & $ $  \\ 
$d^{\prime}$[$\rho_{0}$] & $ 0.004$          & $0.002$           & $ 0.004$          & $0.010$             \\ 
$d^{\prime}$[$\rho_{e}$] & $ 0.005$          & $0.001$           & $ 0.013$          & $0.008$             \\ 
\hline\hline
\end{tabular}
\end{center}
\end{table}

Due to a strong repulsive $\pi^{-}$-nucleus interaction and an attractive Coulomb interaction,
the pions are bound at the surface of nuclei at an effective density $\rho_{e} \sim 0.6\rho_0$ \cite{KIEN}. 
The nontrivial momentum dependence of the Jost function and the Breit-Wigner amplitude results 
in a density dependence of the parameters $B_0$ and $C_0$. Table 1 shows, respectively, two 
sets of the $d^{\prime}$ parameters for $\rho_0$ and $\rho_{e}$. The in-medium $d^{\prime}$ width is 
scaled according to the reduced density.

The $d^{\prime}(2065)$ corrections to the $B_0$ and $C_0$ at $\rho_{e} \sim 0.6\rho_0$ 
where the pion-nucleus interaction has the strongest effect on the pionic atoms are currently
within the empirical errors. If $\pi NN$-coupled resonances exist, the progress in pionic X-ray 
spectroscopy can reveal their contributions to energy levels and widths of the $\pi^{-}$-nucleus bound states.

\vspace{2mm}

The authors are grateful to M. G. Schepkin for useful remarks. M.I.K. and B.V.M. wish to acknowledge kind hospitality at the University of Tuebingen. This work is 
supported by DFG grant No. 436 RUS 113/721/0-1 and RFBR grant No. 03-02-04004.



\begin{thebibliography}{99}
\bibitem{JAFF} R. L. Jaffe, Phys. Rev. Lett. {\bf 38}, 195 (1977); Erratum: Ibid {\bf 38}, 617 (1977).
\bibitem{MULD} P. J. Mulders, A. T. Aerts and J. J. de Swart, Phys. Rev. {\bf D21}, 2653 (1980). 
\bibitem{KON} L. A. Kondratyuk, B. V. Martemyanov and M. G. Schepkin, Sov. J. Nucl. Phys. {\bf 45}, 776 (1987).
\bibitem{FILK} L.V. Fil'kov, V.L. Kashevarov, E.S. Konobeevski, M.V. Mordovskoy, S.I. Potashev, V.M. Skorkin, Phys. Rev. {\bf C61}, 044004 (2000). 
\bibitem{SAKA}  T. Sakai, K. Shimizu, K. Yazaki, Prog. Theor. Phys. Suppl. {\bf 137}, 121 (2000). 
\bibitem{BILG} R. Bilger, H. A. Clement and M. G. Schepkin, Phys. Rev. Lett. {\bf 71}, 42 (1993). 
\bibitem{MASCH} B. V. Martemyanov and M. G. Schepkin, JETP Lett. {\bf 53}, 139 (1991). 
\bibitem{AOKI} S. Aoki et al., Phys. Rev. Lett. {\bf 65}, 1729 (1990).
\bibitem{BARM96} V. V. Barmin et al., Phys. Lett. {\bf B370}, 233 (1996). 
\bibitem{BELZ}  J. Belz et al., Phys. Rev. Lett. {\bf 76}, 3277 (1996); Phys. Rev. {\bf C56}, 1164 (1997).
\bibitem{CHRI98} R. E. Chrien, Nucl. Phys. {\bf A629}, 388c (1998). 
\bibitem{SIOD} U. Siodlaczek et al., Eur. Phys. J. {\bf A9}, 309 (2000). 
\bibitem{DORO} E. Doroshkevich et al., Eur. Phys. J. {\bf A18}, 171 (2003).
\bibitem{GIBB} M. Nuseirat, M.A.K. Lodhi, M.O. El-Ghossain, W.R. Gibbs, W.B. Kaufmann, Phys. Rev. {\bf C58}, 2292 (1998). 
\bibitem{BOBY} Yu-X. Liu, Amand Faessler, J. Schwieger and A. Bobyk, J. Phys. {\bf G24}, 1135 (1998). 
\bibitem{JTPL} M.~I.~Krivoruchenko,
JETP Lett.\  {\bf 46}, 3 (1987).
\bibitem{PRC} A.~Faessler, A.~J.~Buchmann and M.~I.~Krivoruchenko,
Phys.\ Rev.\ {\bf C56}, 1576 (1997).
\bibitem{JPG} A.~Faessler, A.~J.~Buchmann, M.~I.~Krivoruchenko and B.~V.~Martemyanov,
J.\ Phys.\ {\bf G24}, 791 (1998).
\bibitem{GLE} N. K. Glendenning and J. Schaffner-Bielich, Phys. Rev. {\bf C58}, 1298 (1998). 
\bibitem{KONI} J. Koni et al., Nucl. Phys. {\bf A326}, 401 (1979).
\bibitem{OLIN} A. Olin et al., Nucl. Phys. {\bf A439}, 589 (1985).
\bibitem{LAAT} C. T. A. M. de Laat et al., Phys. Lett. {\bf B162}, 81 (1985). 
\bibitem{YAM98} T. Yamazaki et al., Phys. Lett. {\bf B418}, 246 (1998).
\bibitem{GEI02}  H. Geissel et al., Phys. Rev. Lett. {\bf 88}, 122301 (2002).
\bibitem{KIEN} P. Kienle and T. Yamazaki, Prog. Part. Nucl. Phys. {52}, 85 (2004).
\bibitem{SCHE} M. Shchepkin, O. Zaboronsky, H. Clement, Z. Phys. {\bf A345} 407 (1993).
\bibitem{GOLD} M. L. Goldberger and K. M. Watson, {\it Collision Theory}, John Wiley and Sons, NY (1965).
\bibitem{HI} S. M. Kiselev, M. I. Krivoruchenko, B. V. Martemyanov, Amand Faessler and C. Fuchs, Nucl. Phys. {\bf A650}, 78 (1999).
\bibitem{BRJE} G. E. Brown and A. D. Jackson, {\it The Nucleon-Nucleon Interaction}, North-Holland Publishing Company, Amsterdam, 1976.
\bibitem{ERIC} M. Ericson and T. E. O. Ericson, Ann. Phys. (N.Y.) {\bf 36}, 323 (1966).
\bibitem{GARC} C. Garcia-Recio, E. Oset and L. L. Salcedo, Phys. Rev. {\bf C37}, 194 (1988).
\bibitem{OSET} E. Oset, C. Garcia-Reico and J. Nieves, Nucl. Phys. {\bf A584}, 653 (1995).
\bibitem{CMKO} C. M. Ko and D. O. Riska, Nucl. Phys. {\bf A312}, 217 (1978).
\bibitem{SALC} L. L. Salcedo, K. Holinde, E. Oset and C. Schutz, Phys. Lett. {\bf B353}, 1 (1995).
\bibitem{WEIS} W. Weise, Nucl. Phys. {\bf A690}, 98 (2001).
\bibitem{FRIE} E. Friedman, Phys. Lett. {\bf B524}, 87 (2002).
\bibitem{FRIE02} E. Friedman, Nucl. Phys. {\bf A710}, 117 (2002). 
\bibitem{BIRB83} B. L. Birbrair, V. N. Fomenko, A. B. Gridnev and Yu. A. Kalashnikov, J. Phys. {\bf G9}, 1473 (1983); {\bf G11}, 471 (1985). 
\bibitem{GOUD91} P. F. A. Goudsmit, H. J. Leisi and E. Matsinos, Phys. Lett. {\bf B271}, 290 (1991).
\bibitem{BIRB91} B. L. Birbrair and A. B. Gridnev, Nucl. Phys. {\bf A528}, 647 (1991).
\bibitem{AGAL92} A. Gal, B. K. Jennings and E. Friedman, Phys. Lett. {\bf B281}, 11 (1992).
\bibitem{BIRB92} B. L. Birbrair, A. B. Gridnev, L. P. Lapina, A. A. Petrunin and A. I. Smirnov, Nucl. Phys. {\bf A547}, 645 (1992).
\end{thebibliography}
\end{document}